\documentstyle[preprint,revtex]{aps}

\def\ts{\textstyle}

\def\nin{\noindent}

\def\arg{{\rm arg}}

\begin{document}

\begin{flushright}{ UCH-93/ PUC-93/ CECS-93 \\
      {\rm\today} }\end{flushright}

\begin{title}
Bose-Fermi Transformation in Three-Dimensional Space
\end{title}

\author{Luis Huerta}
\begin{instit}
Facultad de F\'\i sica, P. Universidad Cat\'olica de Chile,
Casilla 306, Santiago 22, Chile
\end{instit}
\moreauthors{Jorge Zanelli}
\begin{instit}
Departamento de F\'\i sica, Facultad de Ciencias, Universidad de
Chile,\\ Casilla 653, Santiago, Chile\\ and\\ Centro de Estudios
Cient\'\i ficos de Santiago, Casilla 16443, Santiago 9, Chile
\end{instit}

\begin{abstract}

\nin
A generalization of the Jordan-Wigner transformation to three (or
higher) dimensions is constructed. The nonlocal mapping of spin to
fermionic variables is expressed as a gauge transformation with
topological charge equal to one.  The resulting fermionic theory is
minimally coupled to a nonabelian gauge field in a spontaneously
broken phase containing monopoles.

\end{abstract}


\narrowtext

The Jordan--Wigner (J--W) transformation \cite{jordan28} for
one--dimensional spin systems has provided remarkable applications in
condensed matter physics, including the two--dimensional classical
Ising model \cite{onsager44,kadanoff71} and the $XY$ spin--1/2 model
\cite{schultz64}.  The counterpart in relativistic field theory, the
bosonization of fermionic theories in 1+1 dimensions
\cite{coleman75}, has also opened an important field of active
research.

Bosonization in higher dimensions has been elusive for a long time.
Relatively recent work has uncovered a Bose--Fermi transmutation in
2+1-dimensions which is experienced by the elementary excitations of
the sigma model in the presence of a Chern--Simons field
\cite{polyakov88,huerta89}. This result paved the way for the
construction of the \makebox{J--W} transformation in a lattice of two
spatial dimensions, where a local fermion theory is mapped onto a
system of hard--core bosons described by the Heisenberg Hamiltonian
\cite{fradkin89}.  On the same basis, the bosonization scheme has
been also implemented for 2+1 relativistic field theory
\cite{luscher89}.

In this letter, we propose an extension of the J--W transformation to
three --or more-- dimensions. Here we discuss in detail the
three-dimensional case. The generalization to higher dimensions is
straightforward.

The J--W transformation relates the local spin--1/2 operators,
$S^{\rm z}$,$S^{\pm}$ ($[S^{\rm z},S^{\pm}]=\pm S^{\pm}$,
$[S^+,S^-]=2S^{\rm z}$), to local fermionic operators, $\psi$,
$\psi^{\dag}$ ($\{\psi,\psi^{\dag}\}=1$,
$\{\psi,\psi\}=\{\psi^{\dag},\psi^{\dag}\}=0$):

\begin{equation}
S^+({\bf x}) = \psi^{\dag}({\bf x})\,U({\bf x})\;,
\quad S^-({\bf x}) = U^{\dag}({\bf x})\psi({\bf x}) \;,
\label{eq:jw}
\end{equation}
where $U({\bf x})$ is a nonlocal function of $\psi$.

In a one--dimensional lattice (1D), the operator $U$ takes the
form

\begin{equation}
U_{\rm 1D}({\bf x}) = e^{\ts i\pi\sum_{{\bf z}\neq{\bf x}}
j^0({\bf z})\theta({\bf z}-{\bf x})} \;,
\label{eq:jw1d}
\end{equation}
where $j^0({\bf z})=\psi^{\dag}({\bf z})\psi({\bf z})$ is the fermion
number density operator, and $\theta({\bf z})$ is the 1D step
function.  The corresponding expression in 2D is
\cite{fradkin89,wang91}

\begin{equation}
U_{\rm 2D}({\bf x}) = e^{\ts i\sum_{{\bf z}\neq{\bf x}}
j^0({\bf z})\arg({\bf z}-{\bf x};{\bf\hat{n}}_0)} \;,
\label{eq:jw2d}
\end{equation}
where the function $\arg({\bf x})$ is the angle between ${\bf x}$ and
an arbitrarily given space direction ${\bf\hat{n}}_0$.

The 3D J--W transformation has the same form as (\ref{eq:jw}), but
now it connects an SU(2) doublet of spins $S_{\alpha}$ to an SU(2)
doublet of fermion operators $\psi _{\alpha}$:

\begin{equation}
S^+_{\alpha}({\bf x}) = \psi^{\dag}_{\alpha}({\bf x})
e^{\ts i\sum_{{\bf z}\neq{\bf x}} j^{0a}({\bf z})
\omega^a({\bf z}-{\bf x};{\bf\hat{n}}_0)} \;,
\label{eq:jw3d}
\end{equation}
where $j^{0a}({\bf z})\equiv\psi^{\dag}({\bf z})\tau^a\psi({\bf z})$
is an SU(2) ``isospin'' density operator
($[\tau^a,\tau^b]=i\epsilon^{abc}\tau^c$, sum over repeated indices
is implied), and

\begin{equation}
\omega^a({\bf x};{\bf\hat{n}}_0) = \arg({\bf x};{\bf\hat{n}}_0)
\,e^a({\bf x};{\bf\hat{n}}_0) \;,
\label{eq:omega}
\end{equation}
with $e^a({\bf x};{\bf\hat{n}}_0)$ being a unit vector orthogonal to
${\bf x}$ and ${\bf\hat{n}}_0$.  The application
${\bf x}\to\omega^a({\bf x};{\bf\hat{n}}_0)$ generalizes the 2D and
1D expressions, as it can be seen by restricting it to a plane and to
a line, respectively.

The mapping is completed by exhausting the commutator algebra of
$S^+_{\alpha}$ and $S^-_{\alpha}$. In particular, the generalization
of $S^{\rm z}$, $S^{\rm z}_{\alpha\beta}({\bf x})
\equiv(1/2)[S^+_{\alpha}({\bf x}),S^-_{\beta}({\bf x})]$, is

\begin{equation}
S^{\rm z}_{\alpha\beta} =
{1\over2}[1-\rho({\bf x})]\delta_{\alpha\beta}
-{1\over2}j^{0a}({\bf x})\tau^a_{\alpha\beta} \;,
\end{equation}
where $\rho({\bf x})\equiv\psi^{\dag}_{\alpha}({\bf x})
\psi_{\alpha}({\bf x})$ is the fermion density. It is readily seen
that the diagonal part, $S^{\rm z}_{\alpha\alpha}$, has the usual
form, $1/2-\psi^{\dag}_{\alpha}\psi_{\alpha}$ (no sum over $\alpha$).
The inverse of (\ref{eq:jw3d}) reads

\begin{equation}
\psi^{\dag}_{\alpha}({\bf x}) = S^+_{\alpha}({\bf x})
\,e^{\ts -i\sum_{{\bf z}\neq{\bf x}}S^{{\rm z}a}({\bf z})
\omega^a({\bf z}-{\bf x};{\bf\hat{n}}_0)} \;,
\label{eq:inv}
\end{equation}
with
$S^{{\rm z}a}\equiv -S^{\rm z}_{\alpha\beta}\tau^a_{\beta\alpha}$.

The key feature of the ansatz (\ref{eq:jw3d})--(\ref{eq:omega}),
which is responsible for the transmutation of statistics, is the
fact that

\begin{equation}
\omega^a({\bf y}-{\bf x};{\bf\hat{n}}_0)
- \omega^a({\bf x}-{\bf y};{\bf\hat{n}}_0) =
\pi\,e^a({\bf x}-{\bf y};{\bf\hat{n}}_0) \;.
\label{eq:pi}
\end{equation}
This gives rise to a $(-1)$ factor when the positions of two spins
are exchanged, leading to opposite statistics for the $S$ and $\psi$
operators \cite{notheta}.  For ${\bf x}\neq{\bf y}$, one finds
\cite{rules}

\widetext
\FL
\begin{eqnarray}
S^-_{\alpha}({\bf x}) S^+_{\rho}({\bf y})&&
\biggm[ e^{\ts i\tau^a\omega^a({\bf y}-{\bf x};{\bf\hat{n}}_0)}
\biggm]_{\rho \beta}
-\, S^+_{\beta}({\bf y}) S^-_{\rho}({\bf x} )
\biggm[ e^{\ts i\tau^a\omega^a({\bf y}-{\bf x};{\bf\hat{n}}_0)}
\biggm]_{\alpha \rho} \nonumber\\
= &&\psi_{\alpha}({\bf x})\psi^{\dag}_{\beta}({\bf y})
U^{\dag}({\bf x}) U({\bf y}) \nonumber\\
&&- \psi^{\dag}_{\beta}({\bf y})\psi_{\rho}({\bf x})
\biggm[ e^{\ts i\tau^a\omega^a({\bf y}-{\bf x};{\bf\hat{n}}_0)}
e^{\ts -i\tau^a\omega^a({\bf x}-{\bf y};{\bf\hat{n}}_0)}
\biggm]_{\alpha\rho} U({\bf y}) U^{\dag}({\bf x}) \;.
\nonumber\\
\label{eq:s+s-}
\end{eqnarray}
\narrowtext
\noindent
By virtue of (\ref{eq:pi}), the exponential on the RHS of
(\ref{eq:s+s-}) is $-\delta_{\alpha\rho}$. On the other hand, one may
choose the reference vector
${\bf\hat{n}}_0=({\bf y}-{\bf x})/|{\bf y}-{\bf x}|$, making the
exponentials on the LHS of (\ref{eq:s+s-}) equal to the identity
\cite{n0}.  Also, $U^{\dag}({\bf x})$ and $U({\bf y})$ commute
because the vectors $\,{\bf z}-{\bf y}$, $\,{\bf z}-{\bf x}$ and
${\bf\hat{n}}_0$, for a generic point ${\bf z}$, lie on the same
plane. Hence, for different sites ${\bf x}$ and ${\bf y}$,

\begin{equation}
[S^-_{\alpha}({\bf x}), S^+_{\rho}({\bf y})] =
\{\psi_{\alpha}({\bf x}),\psi^{\dag}_{\beta}({\bf y})\}
U^{\dag}({\bf x})U({\bf y}) = 0 \;.
\label{eq:[ss]}
\end{equation}
The rest of the ${\bf x}\neq{\bf y}$ commutators can be shown to
vanish along similar lines.  On the other hand, the equal--site
commutators define the algebra of the spin operators, which is a
generalized spin--1/2 algebraic structure \cite{algebra}.

The essential feature of the mapping, responsible for the statistical
transmutation, is its topological structure. The operators $U$ in
(\ref{eq:jw1d}) and (\ref{eq:jw2d}) produce local phase
transformations for the field $\psi({\bf x})$, generated by the
charge density $j^0$. The $U$'s rotate the phase of $\psi$ in a
prescribed manner at each site, throughout the entire lattice.  In
1D, the resulting configuration is a kink centered at ${\bf x}$,
where the $\psi$ fields on the left of ${\bf x}$ are flipped with
respect to those on the right. The 2D operator, on the other hand,
produces a vortex centered at ${\bf x}$.

These local assignments are operations generated by $j^0$ in the
corresponding internal symmetry groups of the fermions (${\bf Z}_2$
and U(1), respectively). Although these are gauge transformations,
they cannot be continuously deformed to the identity due to their
nontrivial homotopical character. The J--W transformation belongs to
the homotopy class of winding number one of the gauge group
\cite{coleman77,actor79}.

Indeed, the J--W transformation establishes a one-to-one
correspondence between the points of the boundary of the lattice
(spatial infinity) and the elements of a group manifold. In 1D, the
boundary $\{-\infty,+\infty\}$ is mapped onto ${\bf Z}_2$; for 2D,
the circle at infinity, $S^1_{\infty}$, is mapped onto U(1).  The
existence of these mappings not continuously connected to the
identity is guaranteed because the zeroth and first homotopy groups
of ${\bf Z}_2$ and U(1) ($\pi_0({\rm Z}_2)$ and $\pi_1({\rm U}(1))$,
respectively) are nontrivial.

The generalization of this construction to 3D, then, calls for a
mapping between the boundary of three dimensional space --the sphere
at infinity $S^2_{\infty}$--, and a group manifold ${\cal M}$ with a
nontrivial second homotopy group, ($\pi_2({\cal M})\neq 0$). The
simplest choice is
${\cal M}=S^2\cong {\rm SU(2)/U(1)}\cong{\rm SO(3)/SO(2)}$, and one
is naturally led to consider the SU(2) or SO(3) gauge symmetry
groups, in a spontaneously broken phase \cite{g/h}. [ For higher
dimensions, the requirement is $\pi_{\rm D-1}({\cal M})\neq0$, and
${\cal M}={\rm SO(D)/SO(D-1)}$, leads one to look for spontaneously
broken SO(D) gauge symmetry. ]

In sum, $U({\bf x})$ in the ansatz (\ref{eq:jw3d})--(\ref{eq:omega})
is a gauge transformation in the homotopy class of winding
number one that, acting on a uniform configuration, produces a
``hedgehog'' arrangement centered at ${\bf x}$.

The 3D Jordan--Wigner transformation provides a fermionic
representation for SU(2)--invariant spin systems.  The spin
operators $S^-_{\alpha}$ and $S^+_{\alpha}$ transform as
$S^-_{\alpha}\to T_{\alpha\beta}S^-_{\beta}$,
$S^+_{\alpha}\to T^*_{\;\;\beta\alpha}S^+_{\beta}$,
${\bf T}\in$SU(2). The simplest SU(2)--invariant Hamiltonian
corresponds to the $XY$ model,

\begin{equation}
H = J\sum_{{\bf x},\hat\mu}
[ S^+_{\alpha}({\bf x})S^-_{\alpha}({\bf x}+\hat\mu) +
S^-_{\alpha}({\bf x})S^+_{\alpha}({\bf x}+\hat\mu) ]
\label{eq:XY}
\end{equation}
($\hat\mu$ runs over the unit cell vectors). Applying the mapping
(\ref{eq:jw3d})--(\ref{eq:omega}) one finds that the product
$U({\bf x})U^{\dag}({\bf x}+\hat\mu)$ becomes the link gauge field in
the fermion hopping:

\begin{eqnarray}
U({\bf x})U^{\dag}({\bf x}{\bf+}\hat\mu) =&& e^{\ts
i\sum_{{\bf z}\neq{\bf x}} j^{0a}({\bf z})\omega^a({\bf z}-{\bf x})}
\,e^{\ts -i\sum_{{\bf z}\neq{\bf x}+\hat\mu}
j^{0a}({\bf z})\omega^a({\bf z}-{\bf x}{\bf-}\hat\mu)}
\nonumber\\
\equiv&& e^{\ts i\sum_{{\bf z}}
j^{0a}({\bf z})W^a_{\mu}({\bf z}-{\bf x})} \;.
\label{eq:link}
\end{eqnarray}
This defines the gauge potential $W^a_{\mu}({\bf z})$, which can be
computed in the continuum,

\begin{equation}
W^a_{\mu}({\bf z}-{\bf x}) =
-\epsilon^{\mu ab}{(z-x)^b\over|{\bf z}-{\bf x}|^2},
\qquad({\bf z}\neq{\bf x})  \;.
\end{equation}
We identify $W^a_{\mu}$ as the potential of a monopole
\cite{thooft74}. Thus, the $XY$ Hamiltonian is mapped to a fermionic
model, minimally coupled to an SU(2) nonabelian gauge field:

\begin{equation}
H = J\sum_{{\bf x},\hat\mu} \psi^{\dag}({\bf x})
\,e^{\ts i{\bf A}_{\mu}({\bf x})}\psi({\bf x}+\hat\mu)
+ H_{\rm G} \;,
\label{eq:Hxy}
\end{equation}
where $\psi({\bf x})=(\psi_1({\bf x}),\psi_2({\bf x}))$, and the
SU(2) gauge field takes the form

\begin{equation}
{\bf A}_{\mu}({\bf x})=
\sum_{{\bf z}\neq{\bf x}}j^{0a}({\bf z})W^a_{\mu}({\bf z}-{\bf x})\;.
\label{eq:amu}
\end{equation}

In (\ref{eq:Hxy}), $H_{\rm G}$ represents the Hamiltonian for the
gauge field degrees of freedom. Its exact expression is not important
to us here, so long as it contains SU(2)$\to$U(1) symmetry--breaking
interactions responsible for the presence of monopoles.  The form of
$H_{\rm G}$ depends on the model under consideration and on the
physical significance one assigns to the gauge field \cite{HG}.

One may view the nonabelian gauge field ${\bf A}_{\mu}$ as a
nondynamical artifact needed for the construction of the
\makebox{J--W} mapping. This point of view, however, would not lead
to a local interaction between fermions (${\bf A}_{\mu}$ would be
just a new name for a nonlocal object). Alternatively, one may regard
the ${\bf A}_{\mu}$ as a dynamical field whose classical equations
possess a solution given by (\ref{eq:amu}).  This approach urges us
to consider the SU(2) gauge symmetry as a true invariance of the
physical system. In fact, the SU(2) gauge symmetry is not foreign to
a spin system on the lattice. Any system of localized spins has a
gauge symmetry which reflects the local freedom in the choice of the
spin quantization axis.  This phenomenon has been recently shown to
give rise to a stability enhancement of the AF ordering in the
Hubbard model.  Also ${\bf A}_{\mu}$ is identified, in that model, as
the field of magnonic excitations \cite{huerta93}.

An additional term that could be included in the Hamiltonian is the
analogue of the Ising interaction, $S^{\rm z}_{\alpha\beta}
({\bf x})S^{\rm z}_{\beta\alpha}({\bf x}+\hat\mu)$. This would
generate a quartic nearest-neighbor interaction for the fermions,

\begin{equation}
{1\over2}[1-\rho({\bf x})][1-\rho({\bf x}+\hat\mu)] +
{1\over2}j^{0a}({\bf x})j^{0a}({\bf x}+\hat\mu) \;,
\label{eq:ising}
\end{equation}
This includes, apart from the usual Ising form (the $\rho({\bf x})$
$\times\rho({\bf x}+\hat\mu)$ term), an additional (iso)spin current
density interaction, $j^{0a}({\bf x})j^{0a}({\bf x}+\hat\mu)$.  This
issue will be discussed elsewhere.

Although our work strictly deals with spin systems in the lattice, it
seems likely that the construction can be extended to the context of
a 3+1 relativistic field theory. The operator $U$ in that case may be
related to the monopole creation operators studied by Marino and
Stephany-Ruiz \cite{marino89}.

In the continuum, the statistical transmutation in the presence of
monopoles is not new. Jackiw and Rebbi, Hasenfratz and 't Hooft, and
Goldhaber \cite{jackiw76} have shown that in an SU(2) gauge theory,
isospin degrees of freedom can be converted into spin degrees of
freedom in the field of a magnetic monopole. If the system has
odd-half integer isospin, a change in statistics is induced.  This
seems to be the reason behind the conspicuous presence of topological
structures in the J--W transformations.

We are grateful to M. Kiwi and C. Teitelboim for many helpful
comments and discussions. We also thank A. Gonz\'alez--Arroyo, who
took part in early discussions of this work.  This work was supported
in part by FONDECYT-Chile grant 193.0910/93, by a European
Communities Research contract, and by institutional funding to the
Centro de Estudios Cient\'\i ficos de Santiago provided by SAREC
(Sweden), and a group of Chilean private companies (COPEC, CMPC,
ENERSIS).


\begin{references}
\bibitem{notheta} In contrast with the 2D case, there are no
intermediate possibilities between Bose and Fermi statistics because
the structure constants of the group completely fix the normalization
of the generators.
\bibitem{algebra} The algebra generated by $S^+_{\alpha}$ and
$S^-_{\alpha}$ can be recognized as the $u$(2,q) Lie algebra.
This is not surprising since in 2D the spin algebra was
$su(2,{\rm c})\cong u(1,{\rm q})$.
\bibitem{g/h} Unlike the 1D and 2D cases, in 3D one is forced to
consider a manifold that is not a Lie group. This is so because
$\pi_2(G) \equiv 0$ for any Lie group (see, e. g., \cite{actor79}).
\bibitem{HG} That is, for example, an SU(2) gauge theory coupled to a
Higgs triplet, with a potential
$V(\phi) =\lambda(\phi^a \phi^a -\mu^2/\lambda)^2$, \mbox{i. e.},
the 't Hooft-Polyakov monopole theory (see Ref. \cite{thooft74}).
\end{references}
\end{document}